\documentclass[a4paper,fleqn,usenatbib]{mnras}

\usepackage{mathptmx}

\usepackage[T1]{fontenc}

\usepackage{graphicx}	
\usepackage{amsmath}	
\usepackage{amssymb}	

\newcommand{\Om}[1]{\Omega_\mathrm{#1}}
\newcommand{\rcb}{r_\mathrm{rcb}}
\newcommand{\Knl}{K_{n\ell}}
\newcommand{\bnl}{\beta_{n\ell}}
\newcommand{\numax}{\nu_\mathrm{max}}
\newcommand{\msun}{\mathrm{M}_\odot}
\newcommand{\rsun}{\mathrm{R}_\odot}

\newcommand{\rh}{r_\mathrm{H}}

\makeatletter
\renewcommand\d[1]{\mspace{6mu}\mathrm{d}#1\@ifnextchar\d{\mspace{-3mu}}{}}
\makeatother

\title[A Diagnostic for Red Giant Differential Rotation]{A Diagnostic for Localizing Red Giant Differential Rotation}

\author[H. Klion and E. Quataert]{
Hannah Klion\thanks{E-mail: hklion@berkeley.edu}
and Eliot Quataert
\\
Astronomy and Physics Departments and Theoretical Astrophysics Center, University of California, Berkeley, Berkeley, CA 94720, USA
}

\date{Accepted XXX. Received YYY; in original form ZZZ}

\pubyear{2016}

\begin{document}
\label{firstpage}
\pagerange{\pageref{firstpage}--\pageref{lastpage}}
\maketitle

\begin{abstract}
We present a simple diagnostic that can be used to constrain the location of the differential rotation in red giants with measured mixed mode rotational splittings. Specifically, in red giants with radii $\sim 4\rsun$, the splittings of p-dominated modes (sound wave dominated) relative to those of g-dominated modes (internal gravity wave dominated) are sensitive to how much of the differential rotation resides in the outer convection zone versus the radiative interior of the red giant. An independently measured surface rotation rate significantly aids breaking degeneracies in interpreting the measured splittings. We apply our results to existing observations of red giants, particularly those of Kepler-56, and find that most of the differential rotation resides in the radiative region rather than in the convection zone. This conclusion is consistent with results in the literature from rotational inversions, but our results are insensitive to some of the uncertainties in the inversion process and can be readily applied to large samples of red giants with even a modest number of measured rotational splittings. We argue that differential rotation in the radiative interior strongly suggests that angular momentum transport in red giants is dominated by local fluid instabilities rather than large-scale magnetic stresses.
\end{abstract}




\section{Introduction}

Asteroseismology with the Kepler satellite has allowed measurements of red giant differential rotation \citep{beck:12}.  In red giants, the frequencies of core gravity (g-) modes  and  envelope pressure (p-) modes overlap, producing what are known as mixed modes \citep[e.\@g.\@][]{aizenman:77}. Broadly, these can be classified as either gravity- or pressure-dominated (hereafter g-m and p-m, respectively), depending on where in the star they have greater amplitude.   Differences between the splittings of p-m and g-m modes show that red giant cores rotate faster than their envelopes \citep{beck:12, mosser:12}. However, inversions of mixed mode splittings have difficulty constraining the exact  location of the differential rotation \citep{deheuvels:14, dimauro:16}.

The measured ratios between the core and envelope angular velocities in red giants are generally smaller than  predicted by stellar evolution calculations that employ standard  models of angular momentum transport \citep{cantiello:14}.   Angular momentum transport is thought to arise from fluid instabilities and/or torques due to large-scale magnetic fields. Since these effects are intrinsically multi-dimensional, and the time-scale of a star's evolution is much longer than its hydrodynamic time-scale, stellar evolution codes must parametrize the (magneto-)hydrodynamic effects of rotation \citep[e.g.][]{eggenberger:08, paxton:13}.

Angular momentum can also be transported by non-rotational instabilities such as convection; the rapid convective mixing time-scales suggest efficient angular momentum transport in convection zones. This is typically taken to imply nearly solid body rotation since convection is assumed to act like a viscosity and eliminate angular velocity gradients. This conclusion is, however, at odds with three-dimensional simulations of rotating convection and analytic arguments, which find significant differential rotation in red giant convection zones \citep{brun:09, kissin:15}. From helioseismology, it is also known that the sun's convection zone is not rotating as a solid body, but rather has differential rotation in latitude as well as in radius \citep{brown:89}.

In red giants, g-m mode splittings are predominantly sensitive to core rotation, while p-m mode splittings can be affected by the rotation profile of the entire star \citep[e.\@g.\@][]{beck:12}. Our goal in this paper is to assess how this difference can be used to constrain the location of the differential rotation. Specifically, 
we compare rotational  splittings for two classes of rotation profiles: one where the differential rotation is concentrated just outside of the hydrogen burning shell \citep[e.g.][]{eggenberger:12}, and another where the differential rotation resides in the convective envelope \citep[e.\@g.\@][]{kissin:15}. We provide simple diagnostics that can be used to distinguish between these different models, complementing more detailed studies based on full inversions.

\vspace{-0.5cm}
\section{Methods}
\label{sec:methods}

\subsection{Stellar Evolution and Asteroseismology Calculations}
\label{subsec:evolution_calcs}

The distribution of red giant masses with observed asteroseismic modes is centred near $1.3 \msun$ \citep{mosser:10, kallinger:10}. We therefore take as our representative model a star with solar metallicity and a zero-age main sequence (ZAMS) mass of $1.33\msun$. We use version 8118 of the 1D stellar evolution code {\sc mesa} to evolve it from the pre-main sequence until it has a radius of $10 \rsun$ as a red giant.

Due to the uncertainties in stellar angular momentum transport, a self-consistently evolved rotational profile would not necessarily be accurate. We therefore evolve our representative model without rotation. This also gives more flexibility in the rotational profiles we can study. The majority of the stellar models and rotation profiles we consider reach at most a few per cent of the surface break-up angular velocity. Modifications to the structure caused by centrifugal forces and rotationally-induced mixing are therefore small.

We calculate the frequencies and eigenfunctions of the dipole mixed modes using the adiabatic version of the pulsation code {\sc gyre}, version 4.1 \citep{townsend:13}. The {\sc mesa} and {\sc gyre} inlists necessary to reproduce our calculations will be posted on \url{http://mesastar.org}.

The frequency of maximum power in the asteroseismic spectrum is $\numax$. Since the mode amplitudes are peaked near this value, we only consider modes within two radial orders of $\numax$, that is ones with frequencies in the interval $[\numax - 2 \Delta \nu, \, \numax + 2 \Delta \nu]$. This is the typical frequency range of observed modes \citep[e.g.\@][]{beck:12, dimauro:16}. $\Delta \nu$ is the large separation, the separation in frequency between p-m modes near $\numax$. We estimate $\Delta \nu$ and $\numax$ for a given stellar model using the scaling relations calibrated in \citet{mosser:10}.

\vspace{-0.5cm}
\subsection{Model Rotation Profiles}
\label{subsec:rot_prof}

We consider an idealized rotation profile in which there is differential rotation both near the core and in the convection zone. Differential rotation  outside  the core is natural if angular momentum transport is dominated by local fluid instabilities that must overcome the strongly stabilizing composition gradient at the hydrogen burning shell. By contrast, if transport is dominated by large-scale magnetic fields, solid-body rotation across the composition gradient is plausible and the star's differential rotation may instead be largely contained within the convection zone.

We assume that the inner portion of the star undergoes solid body rotation at the rate $\Om{c}$. Some of the  differential rotation is concentrated at 1.5 times $\rh$, the outer radius of the hydrogen burning shell. We model this as a step function decrease from $\Om{c}$ to $\Om{m}$, the angular velocity of the middle of the star. Models with a continuous transition between $\Om{c}$ and $\Om{m}$ at $1.5 \rh$ (e.g.\@ a hyperbolic tangent) yield essentially identical results. The remainder of the star's differential rotation is located in the convection zone, where it follows a power law profile. The full profile can therefore be written as

\begin{equation}
  \label{eq:profile}
\Omega(r) =
\begin{cases}
\Omega_\mathrm{c} & r \leq 1.5 \rh \\
\Omega_\mathrm{m} & 1.5 \rh < r \leq \rcb \\
\Omega_\mathrm{e} \left(\frac{R}{r}\right)^\alpha & r > \rcb\\
\end{cases},
\end{equation}
where
\begin{equation}
  \label{eq:alpha}
\alpha = \frac{\log (\Om{m} / \Om{e})}{\log(R/\rcb)}
\end{equation}
is chosen so that $\Omega(\rcb) = \Om{m}$ and $\Omega(R) = \Om{e}$, where $\Om{e}$ is the surface angular velocity and $\rcb$ is the radius of the radiative-convective boundary.

\begin{figure}
   \centering
   \includegraphics{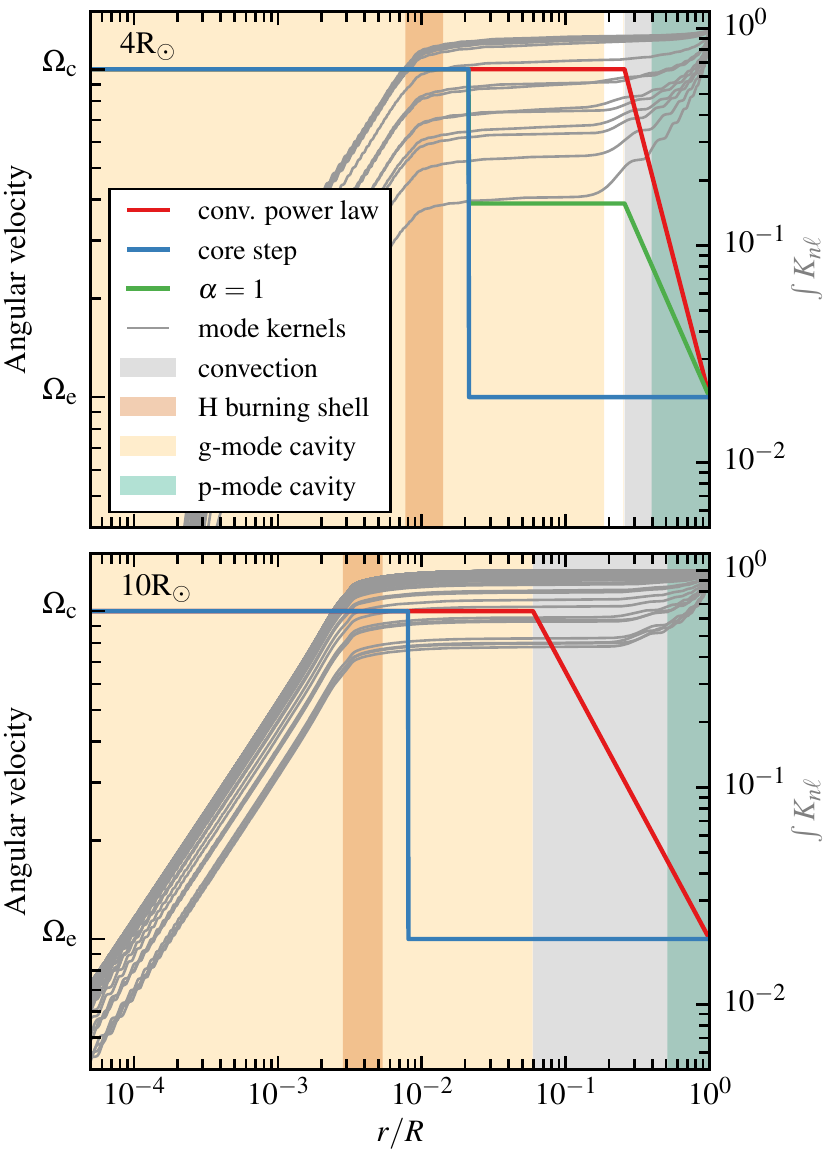}
   \caption{Integrated rotational kernels (grey lines, right y-axes, see equation~(\ref{eq:split})) for asteroseismic modes within two radial orders of $\numax$ calculated using {\sc gyre} for $1.33 \msun$, $4 \rsun$ (top) and $10 \rsun$ (bottom) red giant models evolved using {\sc mesa}. The x-axes show the normalized radial coordinate of the stars. The left y-axes correspond to the coloured lines, which show three of our model rotational profiles: differential rotation in the convection zone (red), near the hydrogen burning shell (blue), or both (green). The hydrogen burning region, g-mode, and p-mode cavities are shaded in light orange, dark orange, and green, respectively. Convection zones are shaded in grey.}
   \label{fig:k_omega}
\end{figure}

We initially focus on two limiting cases. We call the first `convection power law.' All of the differential rotation in this model is contained in the convection zone ($\Om{m} = \Om{c}$ in equation~\ref{eq:profile}). We will be most interested in the $4\rsun$ model, for which $\Om{c}/\Om{e} \sim \hbox{5--30}$ implies $\alpha \sim \hbox{1.2--2.5}$. By comparison, \citet{kissin:15} argue on theoretical grounds that $\alpha \sim \hbox{1--3/2}$. In our alternative `core step' model, all of the  differential rotation is concentrated at $1.5 \rh$. In this case, $\Om{m} = \Om{e}$ and $\alpha = 0$. Examples of these profiles for the $4\rsun$ and $10\rsun$ models are shown in Fig.~\ref{fig:k_omega}. We also show a profile with $\alpha = 1$ for the $4\rsun$ model.

\vspace{-0.5cm}
\subsection{Calculating Mode Frequency Splittings}
\label{subsec:calc_split}

In a non-rotating star, asteroseismic modes with the same spherical harmonic $\ell$ but different $m$ are degenerate in frequency. Rotation introduces a perturbation to the frequency spectrum that causes the modes to split into $(2\ell+1)$-plets at frequencies
\begin{equation}
  \label{eq:split}
\nu_{n\ell m} = \nu_{n\ell} + \delta \nu_{n\ell m} = \nu_{n\ell} + m \frac{\bnl}{2\pi} \int_0^R \Knl(r) \Omega(r) \d r
\end{equation}
where $\nu_{n\ell}$ is the frequency of the unsplit mode. $\bnl$ and $\Knl(r)$ depend on the mode eigenfunctions \citep{christensendalsgaard:14}.  $\Knl$, the rotational kernel, which is normalized to 1 integrated over the star, is greater where the mode has greater amplitude. $\bnl \approx 1/2$ for g-m $\ell = 1$ modes, and approaches $1$ for p-m modes.

To maximize the splitting of a g-m mode, the mode should have nearly all of its amplitude within the core such that its rotational kernel integrated in the core is nearly $1$. These modes also have $\bnl \approx 1/2$. Therefore the theoretical maximum g-m mode splitting is $\max(\delta\nu_\mathrm{g}) = \Om{c}/(4\pi)$ for dipole modes, and we can estimate
\begin{equation}
  \label{eq:om_c}
  \Om{c} \approx 4\pi \max(\delta\nu_\mathrm{g}).
\end{equation}
Empirically, the mode with the largest rotational splitting is a g-m mode, so the maximal splitting $\max(\delta\nu) = \max(\delta\nu_\mathrm{g})$.

The integrated rotational kernels for the $4\rsun$ and $10\rsun$ models (Fig.~\ref{fig:k_omega}) are dominated by contributions from the helium core, hydrogen burning shell, and the outer convection zone, with little contribution from the non-burning radiative hydrogen region. The splittings are insensitive to changes in the location of the differential rotation within the radiative hydrogen region. Our profile with differential rotation concentrated near the hydrogen burning shell is therefore representative of all rotation profiles whose differential rotation is contained within the radiative hydrogen region.

\vspace{-0.5cm}
\section{Results}
\label{sec:results}
\vspace{-0.1cm}

\begin{figure*}
   \centering
   \includegraphics{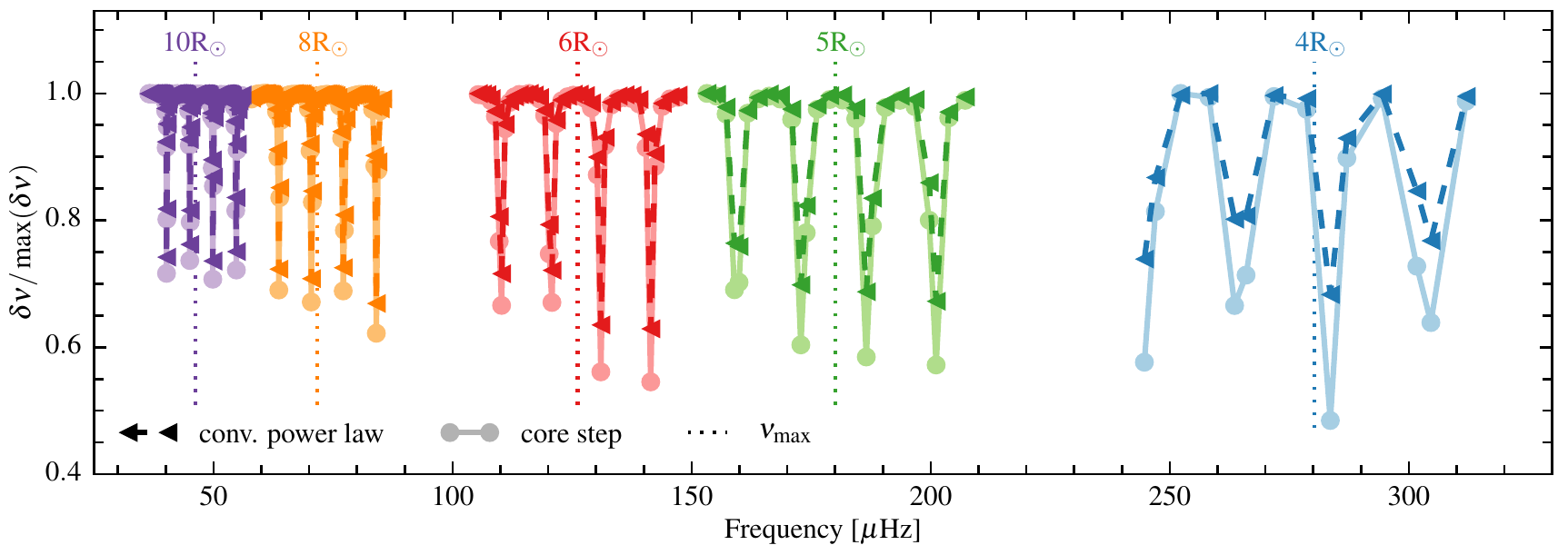} 
   \caption{Rotational mode splittings for models with differential rotation just outside of the hydrogen burning shell (core step, lighter solid lines) and in the convection zone (convection power law, darker dashed lines) at several points along the evolution of a $1.33 \msun$ ZAMS red giant. At each radius, we take $\Om{c} / \Om{e} = 7.89$, and normalize the splittings to the largest splitting at that radius. Each colour denotes a different radius. The dotted lines show $\numax$ for each model. Only splittings for modes within two radial orders of $\numax$ are shown. More evolved stars have smaller values of both $\numax$ and $\Delta \nu$. In the less evolved models ($4\rsun$) the most pressure-dominated p-m modes have splittings that are $\sim \hbox{50--60}$ per cent of the maximum splitting, while more evolved models have $\min(\delta \nu/\max(\delta\nu)) \approx 70$ per cent. The different internal rotation profiles also become less distinguishable at larger radius, i.e.\@ later in the star's evolution.}
 \label{fig:all_splittings}
\end{figure*}

Fig.~\ref{fig:all_splittings} shows frequency splittings for red giants with $R=\hbox{4--10}\rsun$ for modes within two radial orders of $\numax$. These results are for rotation profiles with $\Om{c} / \Om{e} = 7.89$. The splittings are normalized to the maximum splitting for each radius (i.e. $\delta\nu/\max(\delta\nu))$. As the star expands, its effective temperature decreases and its luminosity increases, causing $\numax$ and $\Delta \nu$ to decrease.

The p-m splittings are smaller than the g-m splittings in all models because the surface rotation rate is significantly less than the core rotation rate in these rotation profiles. The ratio between the p-m and g-m splittings increases from $\sim \hbox{50--60}$ per cent in the $4\rsun$ model to $\sim 70$ per cent in the $10 \rsun$ model, indicating that the p-m modes have greater core amplitudes in models with larger radii. In general, higher frequency modes are more p-like. Since $\numax$ is smaller in larger models, modes near $\numax$ will become less p-like as the star evolves, leading to the smaller difference between the p-m and g-m splittings seen in Fig.~\ref{fig:all_splittings} for larger red giants.

For all models shown in Fig.~\ref{fig:all_splittings}, normalized p-m mode splittings are lower when the radiative hydrogen zone (`core step'), instead of the convective envelope (`conv.\@ power law'), is differentially rotating. Concentrating the differential rotation outside of the hydrogen shell produces a smaller p-m/g-m splitting ratio than nearly any other possible rotation profile. Since there is solid body rotation at $\Om{c}$ and $\Om{e}$ in the core and envelope, the g-m and p-m mode splittings are extremized, and their ratio is minimized.

Fig.~\ref{fig:all_splittings} shows that the effect of different rotation profiles on the relative p-m and g-m splittings becomes significantly weaker as the star evolves. Since the differential rotation in both profiles shown in Fig.~\ref{fig:all_splittings} is outside of the core, the core contribution to the splitting is the same in both profiles. The difference in the splittings arises from the difference in the rotation profile in the convective envelope. In red giants with larger radii, the rotation in the convective envelope contributes less to the overall splittings. As a result, the signature of the location of differential rotation is stronger in less evolved red giants, as noted in \citet{goupil:13} and \citet{deheuvels:14}.

\begin{figure}
\centering
\includegraphics{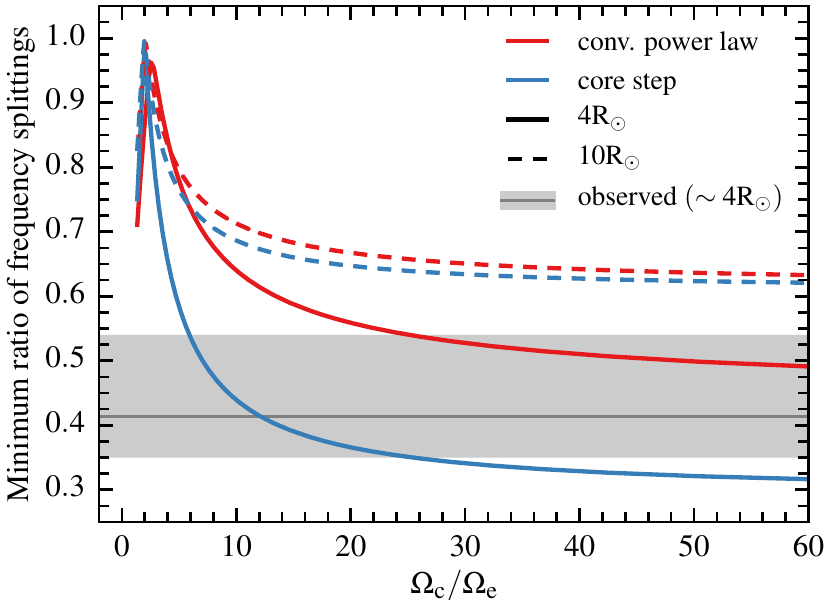} 
\caption{Comparison of minimum ratios of rotational splittings near $\numax$ for models with differential rotation in the convection zone (red, conv.\@ power law) and outside of the hydrogen burning shell (blue, core step) for a range of ratios of core to surface rotation rates, $\Om{c}/\Om{e}$. For $\Om{c}/\Om{e} \gtrsim \hbox{2--3}$ the y-axis represents the smallest value of the ratio of p-m to g-m rotational splittings ($\min(\delta\nu/\max(\delta\nu))$, as in, e.g.\@ Fig.~\ref{fig:all_splittings}). Values for the $4\rsun$ ($10\rsun$) model are shown as solid (dashed) lines. The grey line (shaded region) marks a typical observed value (full range) of $\min(\delta\nu/\max(\delta\nu))$, calculated for five observed $\hbox{1.02--1.49} \msun$ red giants with radii in the range $\hbox{3.79--5.13} \rsun$ (see text). Differential rotation in the convection zone is possible for red giants with larger measured $\min(\delta\nu/\max(\delta\nu))$, but splittings for the majority of $\sim 4 R_\odot$ red giants  are consistent with differential rotation concentrated outside of the core. In general, there is a degeneracy between the location of the differential rotation and $\Om{c}/\Om{e}$, such that the location of the differential rotation can be more robustly constrained if there is an independent measurement of the surface rotation rate, $\Om{e}$.}
   \label{fig:rot_split_ratio}
\end{figure}

Fig.~\ref{fig:rot_split_ratio} shows the minimum normalized splitting for both of our model rotation profiles for a range of core to envelope rotation ratios $\Om{c}/\Om{e}$. The minimum normalized splitting is the smallest value of $\delta\nu/\max(\delta \nu)$ for the range of modes considered (near $\numax$, see Fig.~\ref{fig:all_splittings}). The curves in Fig.~\ref{fig:rot_split_ratio} all have a two-component structure. Below $\Om{c}/\Om{e} \sim \hbox{2--3}$, increasing core rotation relative to the envelope causes the minimum normalized splitting to increase, whereas for faster core rotation the opposite occurs. This is because below $\Om{c}/\Om{e} \sim \hbox{2--3}$, p-m splittings exceed g-m splittings due to the p-m modes' greater values of $\bnl$ (see eq.~\ref{eq:split}). The vast majority of published red giant asteroseismic observations find g-m splittings that exceed p-m splittings, so we focus on profiles with $\Om{c}/\Om{e} \gtrsim 2$.

As expected, Fig.~\ref{fig:rot_split_ratio} shows that the model with differential rotation outside of the hydrogen burning shell has lower minimum normalized splittings than the model with differential rotation in the convection zone. Consistent with  Fig.~\ref{fig:all_splittings}, the $4\rsun$ model has lower minimum normalized splittings and shows a larger difference between the two types of rotation profiles than the $10\rsun$ model. Fig.~\ref{fig:rot_split_ratio} also shows a typical observed value of $\min(\delta\nu/\max(\delta\nu))$, calculated from measured splittings of the red giants KIC 5356201 (0.38), KIC 8366239 (0.54), KIC 12008916 (0.35) \citep{beck:12}, Kepler-56 (0.39) \citep{huber:13}, and KIC 4448777 (0.41) \citep{dimauro:16}. These stars have masses between $\hbox{1.02--1.49}\msun$ and radii in the range $\hbox{3.79--5.13} \rsun$.

Fig.~\ref{fig:rot_split_ratio} highlights an important degeneracy between the location of the differential rotation and $\Om{c}/\Om{e}$. For a given measured minimum normalized splitting, there are in general two possible solutions: a smaller value of $\Om{c}/\Om{e}$ with differential rotation in the core (blue line, core step) or a larger value of $\Om{c}/\Om{e}$ with differential rotation in the convection zone (red line, convection power law). For instance, Fig.~\ref{fig:rot_split_ratio} shows that core differential rotation is generally preferred if $\Om{c}/\Om{e} \sim \hbox{7--20}$ (similar to the conclusion of \citealt{dimauro:16}) but convection zone differential rotation with a more slowly rotating envelope could also be consistent with the observed splittings of some red giants. Note, however, that $\min(\delta\nu/\max(\delta\nu))$ tends to a constant value as $\Om{c}/\Om{e} \rightarrow \infty$. In some cases, this asymptotic value for the profile with differential rotation in the convection zone is so large that this profile is strongly disfavoured (Fig.~\ref{fig:rot_split_ratio}).

The degeneracy between high $\Om{c}/\Om{e} \sim 50$ and lower $\Om{c}/\Om{e} \sim 10$ can be broken if there are enough measured rotational splittings to perform a full inversion. In practice, however, this has been difficult given the finite numbers of available modes \citep{deheuvels:14, dimauro:16}. Consequently, a measurement of $\Om{c}/\Om{e}$ is very useful for estimating the radial location of the differential rotation, particularly given uncertainties in the stellar parameters and structure which further complicate performing full inversions. $\Om{c}$ can be estimated from the splittings of the most core-dominated g-m modes using equation~(\ref{eq:om_c}), but an independent measurement of the surface rotation rate, $\Om{e}$, can significantly help pinpoint the interior location of the differential rotation.

\vspace{-0.5cm}
\section{Kepler-56}
\label{sec:k56}

\citet{huber:13} measured  the dipole mixed modes of Kepler-56, a red giant with a radius of $4.23 \pm 0.15 \rsun$ and a mass of $1.32 \pm 0.13 \msun$. They also found a rotational period of $74 \pm 3$~days, measured from flux variations assumed to be due to starspots. This is nominally more precise than the rotational period of $\hbox{60--230}$~days inferred from  rotational broadening, $v\sin i = 1.7 \pm 1.0 \text{ km s}^{-1}$, where $v$ is the surface rotational velocity, and $i = 47 \pm 6^{\circ}$ is the inclination of the star's rotation axis relative to the line of sight, determined from  asteroseismology. 

\begin{figure}
   \centering
   \includegraphics{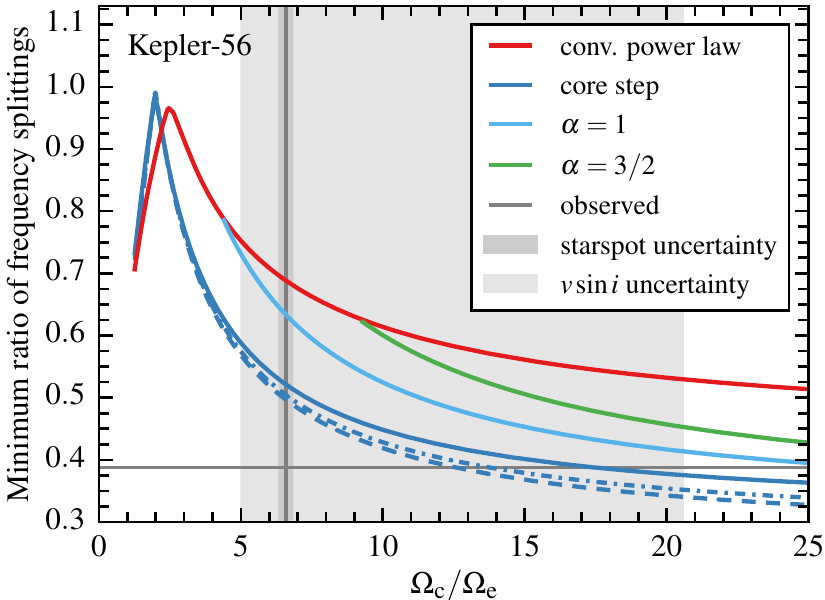}
   \caption{Minimum ratios of rotational splittings (as in Fig.~\ref{fig:rot_split_ratio}) for models of the red giant Kepler-56. Results are shown for a range of $\Om{c}/\Om{e}$, the ratio of the core to surface rotation rates and four rotation profiles with: differential rotation in the convection zone (conv.\@ power law, red), just outside of the hydrogen burning shell (core step, blue), and in both the convection and radiative zones ($\alpha = 1$ and $\alpha = 3/2$, light blue and green, respectively). In these latter models, $\Omega \propto r^{-\alpha}$ in the convection zone (see equation~(\ref{eq:profile})). The observed minimum normalized splitting for Kepler-56 is marked with a grey horizontal line. The darker (lighter) shaded region shows the uncertainty in the measurement of $\Om{c}/\Om{e}$ from flux variations due to starspots (rotational line broadening). The value of $\Om{c}/\Om{e}$ implied by the minimum normalized splitting is inconsistent with the starspot measurement, but consistent with spectroscopic estimates of the surface rotation rate. The convection power law and $\alpha = 3/2$ models (red and green lines) are inconsistent with the data implying that most of the differential rotation cannot be in the convection zone, but must instead primarily reside in the core. $4.08$ (dashed), $4.23$ (solid), and $4.38\rsun$ (dashed-dotted line) models bracket the measured radii and have little effect on our conclusions. }
   \label{fig:k56}
\end{figure}

Fig.~\ref{fig:k56} shows the minimum ratios of rotational splittings for a model with radius $4.23\rsun$. \citet{huber:13} only report rotational splittings for select modes, so we estimate the remaining splittings as the average splitting of the $m=\pm 1$ modes. For the rotation profile with differential rotation near the core, we also show results for models with radii of $4.08\rsun$ and $4.38\rsun$ corresponding to the observed radius plus or minus the uncertainty. In addition to our two default rotation profiles, we consider two additional rotation profiles with differential rotation both near the core and in the convection zone. For these profiles, we fix $\alpha$ (see equations~(\ref{eq:profile}) and (\ref{eq:alpha})) at 1 and 3/2, bracketing the range of $\alpha$ considered in \citet{kissin:15}. We only consider profiles with $\Om{c} \geq \Om{m} \geq \Om{e}$, which means that the models with fixed $\alpha$ require $\Om{c}/\Om{e} \geq 4.4$ and $9.1$ for $\alpha = 1$ and $3/2$, respectively.

In Kepler-56, the minimum observed ratio between g-m and p-m splittings is $0.482 \,\mu\mathrm{Hz} / 0.198 \,\mu\mathrm{Hz} = 0.388$ and is marked as a grey horizontal line in Fig.~\ref{fig:k56}. We estimate $\Om{c}$ using equation~(\ref{eq:om_c}) and compute $\Om{c}/\Om{e}$ and its uncertainty using both measurements of $\Om{e}$ (see Fig.~\ref{fig:k56}). Fig.~\ref{fig:k56} strongly suggests that differential rotation predominantly in the convection zone (`conv.\@ power law') is disfavoured.  Fig.~\ref{fig:k56} also shows that none of the rotation profiles we consider is consistent with the starspot-inferred $\Om{c}/\Om{e}$ and observed $\min(\delta\nu/\max(\delta\nu))$. The splittings are, however, consistent with the value of $\Om{c}/\Om{e}$ inferred from $v\sin i$ if the differential rotation is primarily near the core (`core step'). The $v \sin i$ measurement and observed normalized splittings are also consistent with $\alpha \sim 1$, which corresponds to a factor of $\sim 4$ change in angular velocity across the convection zone. 

The minimum ratio between the g-m and p-m splittings is lowest when the differential rotation is in the non-burning hydrogen region, where the rotational kernels are very small. No other rotation profile will have substantially lower minimum normalized splittings than the core step profile (dark blue line in Fig.~\ref{fig:k56}). The calculated splittings are also largely insensitive to details of the stellar model. Fig.~\ref{fig:k56} shows that the model radius has only a small effect on the normalized splittings. We have also varied the mixing length parameter, the stellar mass, and the metallicity. None of these changes significantly affects the minimum splitting ratios shown in Fig.~\ref{fig:k56}. In all cases, the minimum splitting ratio in Kepler-56 implies that $\Om{c}/\Om{e}$ is $\hbox{2--3}$ times the value given by the starspot measurements. Since the core rotation rate is well-known from the maximum g-m splittings, this suggests that the starspot measurement of the surface rotation rate is too large by a factor of $\sim 2$.

\vspace{-0.5cm}
\section{Discussion}
\label{sec:discussion}

Measurements of the rotational splittings of mixed modes in red giants show that the core is rotating much faster than the envelope but have not definitively determined where in the interior most of the differential rotation is localized \citep{beck:12, deheuvels:14, dimauro:16}. Constraining the location of the differential rotation within the interior would strongly constrain the dominant mechanisms of angular momentum transport in red giants. We have shown that the ratio between the minimum p-m rotational splitting and the maximum g-m splitting is a simple yet effective diagnostic for localizing the differential rotation in the interiors of red giants (Figs.~\ref{fig:all_splittings} and \ref{fig:rot_split_ratio}). It has the most discriminating power for modestly evolved red giants with radii $\sim 4\rsun$. A non-asteroseismic measurement of the surface rotation is particularly valuable in applying this diagnostic because it anchors the contribution of the surface rotation to the splittings of the p-m modes (Fig.~\ref{fig:k56}). The technique considered here allows for the rapid characterization of differential rotation in large numbers of red giants without the need for many measured frequency splittings or detailed rotational inversions, which are uncertain because of the small numbers of modes available in red giants.

We find that for a typical $\sim \hbox{1--1.5}\msun$ red giant with a radius of $\sim \hbox{4--5}\rsun$, observed frequency splittings are best explained by differential rotation just outside of the hydrogen burning shell and a core that rotates $\sim {5–-20}$ times faster than the envelope (Fig.~\ref{fig:rot_split_ratio}). This is consistent with prior studies that have calculated red giant rotation profiles by inverting the frequency splittings \citep[e.g.][]{deheuvels:14, dimauro:16}. A much more slowly rotating envelope with substantial differential rotation in the convection zone is disfavoured in general but can produce measured p-m to g-m splitting ratios at the upper end of the measured values (Fig.~\ref{fig:rot_split_ratio}). Surface rotation measurements would break the degeneracy between $\Om{c}/\Om{e}$ and the location of the differential rotation.

Kepler-56 has estimated surface rotation rates from both rotational line broadening and spots. As Fig.~\ref{fig:k56} shows, the measured $v \sin i$ rules out most of the differential rotational residing in the convection zone. Instead, differential rotation near the H burning shell with $\Om{c}/\Om{e} \sim \hbox{13--17}$ is preferred. Alternatively, differential rotation can exist both near the core and in the convection zone, but the net change in rotation rate across the convection zone is constrained to be less than a factor of $\sim 4$ (corresponding to alpha $\lesssim 1$ in Fig.~\ref{fig:k56}). Our inferred surface rotation period for Kepler-56 suggests that the starspot measurement of the surface rotation rate from \citet{huber:13} is too large by a factor of $\sim 2$. This is plausible because long-time-scale flux variations (including any due to rotation) are removed in calibrating and de-trending Kepler lightcurves.

\citet{kissin:15} proposed that the Kepler core rotation measurements of red giants could be explained by differential rotation residing primarily in the convection zone rather than in the radiative interior. They showed that the evolution of red giant core rotation rates with increasing radius is broadly consistent with theoretical models in which the rotation rate varies as $\Omega \sim r^{-\alpha}$ in convection zones, with $\alpha \sim \hbox{1--3/2}$. Significant differential rotation in convection zones is also consistent with numerical simulations \citep{brun:09} and theoretical arguments based on geostrophic balance and/or conservation of angular momentum by convective eddies \citep{kissin:15}. Our results strongly disfavour \citet{kissin:15}'s proposal that {\em most} of the differential rotation in red giants is in the convection zone (Fig.~\ref{fig:rot_split_ratio} and \ref{fig:k56}). However, models with some differential rotation in the convection zone ($\alpha \sim 1$) are marginally consistent with the measured rotational splittings and the surface rotation rate of Kepler-56. Constraining better the degree of differential rotation in red giant convection zones would be very valuable because this is one of the significant uncertainties in models of rotating stellar evolution. Doing so would require a larger sample of accurate surface rotation rates for $R \sim 4 \rsun$ red giants with seismology, either from spectroscopy or spot measurements.

The presence of strong differential rotation in the non-burning hydrogen region of Kepler-56 (and other red giants) strongly suggests that angular momentum transport in the interior is dominated by small-scale instabilities that cannot overcome the strongly stabilizing composition gradient at the edge of the helium core. By contrast, large scale magnetic fields are disfavoured because they would be expected to maintain solid body rotation across this stabilizing composition gradient.

\vspace{-0.5cm}

\section*{Acknowledgements}
We thank Lars Bildsten, Gibor Basri, and Jim Fuller for useful conversations. HK is supported by a DOE Computational Science Graduate Fellowship under grant number DE-FG02-97ER25308. EQ is supported in part by a Simons Investigator award from the Simons Foundation, the David and Lucile Packard Foundation, and the Gordon and Betty Moore Foundation through Grant GBMF5076. The simulations presented here were carried out and processed using the Savio computational cluster resource provided by the Berkeley Research Computing program at the University of California, Berkeley (supported by the UC Berkeley Chancellor, Vice Chancellor of Research, and Office of the CIO).




\bibliographystyle{mnras}
\bibliography{bibliography} 

\begin{thebibliography}{}
\makeatletter
\relax
\def\mn@urlcharsother{\let\do\@makeother \do\$\do\&\do\#\do\^\do\_\do\%\do\~}
\def\mn@doi{\begingroup\mn@urlcharsother \@ifnextchar [ {\mn@doi@}
  {\mn@doi@[]}}
\def\mn@doi@[#1]#2{\def\@tempa{#1}\ifx\@tempa\@empty \href
  {http://dx.doi.org/#2} {doi:#2}\else \href {http://dx.doi.org/#2} {#1}\fi
  \endgroup}
\def\mn@eprint#1#2{\mn@eprint@#1:#2::\@nil}
\def\mn@eprint@arXiv#1{\href {http://arxiv.org/abs/#1} {{\tt arXiv:#1}}}
\def\mn@eprint@dblp#1{\href {http://dblp.uni-trier.de/rec/bibtex/#1.xml}
  {dblp:#1}}
\def\mn@eprint@#1:#2:#3:#4\@nil{\def\@tempa {#1}\def\@tempb {#2}\def\@tempc
  {#3}\ifx \@tempc \@empty \let \@tempc \@tempb \let \@tempb \@tempa \fi \ifx
  \@tempb \@empty \def\@tempb {arXiv}\fi \@ifundefined
  {mn@eprint@\@tempb}{\@tempb:\@tempc}{\expandafter \expandafter \csname
  mn@eprint@\@tempb\endcsname \expandafter{\@tempc}}}

\bibitem[\protect\citeauthoryear{{Aizenman}, {Smeyers}  \&
  {Weigert}}{{Aizenman} et~al.}{1977}]{aizenman:77}
{Aizenman} M.,  {Smeyers} P.,   {Weigert} A.,  1977, \aap, \href
  {http://adsabs.harvard.edu/abs/1977A%26A....58...41A} {58, 41}

\bibitem[\protect\citeauthoryear{{Beck} et~al.,}{{Beck} et~al.}{2012}]{beck:12}
{Beck} P.~G.,  et~al., 2012, \mn@doi [\nat] {10.1038/nature10612}, \href
  {http://adsabs.harvard.edu/abs/2012Natur.481...55B} {481, 55}

\bibitem[\protect\citeauthoryear{{Brown}, {Christensen-Dalsgaard},
  {Dziembowski}, {Goode}, {Gough}  \& {Morrow}}{{Brown}
  et~al.}{1989}]{brown:89}
{Brown} T.~M.,  {Christensen-Dalsgaard} J.,  {Dziembowski} W.~A.,  {Goode} P.,
  {Gough} D.~O.,   {Morrow} C.~A.,  1989, \mn@doi [\apj] {10.1086/167727},
  \href {http://adsabs.harvard.edu/abs/1989ApJ...343..526B} {343, 526}

\bibitem[\protect\citeauthoryear{{Brun} \& {Palacios}}{{Brun} \&
  {Palacios}}{2009}]{brun:09}
{Brun} A.~S.,  {Palacios} A.,  2009, \mn@doi [\apj]
  {10.1088/0004-637X/702/2/1078}, \href
  {http://adsabs.harvard.edu/abs/2009ApJ...702.1078B} {702, 1078}

\bibitem[\protect\citeauthoryear{{Cantiello}, {Mankovich}, {Bildsten},
  {Christensen-Dalsgaard}  \& {Paxton}}{{Cantiello}
  et~al.}{2014}]{cantiello:14}
{Cantiello} M.,  {Mankovich} C.,  {Bildsten} L.,  {Christensen-Dalsgaard} J.,
  {Paxton} B.,  2014, \mn@doi [\apj] {10.1088/0004-637X/788/1/93}, \href
  {http://adsabs.harvard.edu/abs/2014ApJ...788...93C} {788, 93}

\bibitem[\protect\citeauthoryear{Christensen-Dalsgaard}{Christensen-Dalsgaard}{2014}]{christensendalsgaard:14}
Christensen-Dalsgaard J.,  2014, {Lecture Notes on Stellar Oscillations},
  \url{http://astro.phys.au.dk/~jcd/oscilnotes/}

\bibitem[\protect\citeauthoryear{{Deheuvels} et~al.,}{{Deheuvels}
  et~al.}{2014}]{deheuvels:14}
{Deheuvels} S.,  et~al., 2014, \mn@doi [\aap] {10.1051/0004-6361/201322779},
  \href {http://adsabs.harvard.edu/abs/2014A%26A...564A..27D} {564, A27}

\bibitem[\protect\citeauthoryear{{Di Mauro} et~al.,}{{Di Mauro}
  et~al.}{2016}]{dimauro:16}
{Di Mauro} M.~P.,  et~al., 2016, \mn@doi [\apj] {10.3847/0004-637X/817/1/65},
  \href {http://adsabs.harvard.edu/abs/2016ApJ...817...65D} {817, 65}

\bibitem[\protect\citeauthoryear{{Eggenberger}, {Meynet}, {Maeder}, {Hirschi},
  {Charbonnel}, {Talon}  \& {Ekstr{\"o}m}}{{Eggenberger}
  et~al.}{2008}]{eggenberger:08}
{Eggenberger} P.,  {Meynet} G.,  {Maeder} A.,  {Hirschi} R.,  {Charbonnel} C.,
  {Talon} S.,   {Ekstr{\"o}m} S.,  2008, \mn@doi [\apss]
  {10.1007/s10509-007-9511-y}, \href
  {http://adsabs.harvard.edu/abs/2008Ap%26SS.316...43E} {316, 43}

\bibitem[\protect\citeauthoryear{{Eggenberger}, {Montalb{\'a}n}  \&
  {Miglio}}{{Eggenberger} et~al.}{2012}]{eggenberger:12}
{Eggenberger} P.,  {Montalb{\'a}n} J.,   {Miglio} A.,  2012, \mn@doi [\aap]
  {10.1051/0004-6361/201219729}, \href
  {http://adsabs.harvard.edu/abs/2012A%26A...544L...4E} {544, L4}

\bibitem[\protect\citeauthoryear{{Goupil}, {Mosser}, {Marques}, {Ouazzani},
  {Belkacem}, {Lebreton}  \& {Samadi}}{{Goupil} et~al.}{2013}]{goupil:13}
{Goupil} M.~J.,  {Mosser} B.,  {Marques} J.~P.,  {Ouazzani} R.~M.,  {Belkacem}
  K.,  {Lebreton} Y.,   {Samadi} R.,  2013, \mn@doi [\aap]
  {10.1051/0004-6361/201220266}, \href
  {http://adsabs.harvard.edu/abs/2013A%26A...549A..75G} {549, A75}

\bibitem[\protect\citeauthoryear{{Huber} et~al.,}{{Huber}
  et~al.}{2013}]{huber:13}
{Huber} D.,  et~al., 2013, \mn@doi [Science] {10.1126/science.1242066}, \href
  {http://adsabs.harvard.edu/abs/2013Sci...342..331H} {342, 331}

\bibitem[\protect\citeauthoryear{{Kallinger} et~al.,}{{Kallinger}
  et~al.}{2010}]{kallinger:10}
{Kallinger} T.,  et~al., 2010, \mn@doi [\aap] {10.1051/0004-6361/201015263},
  \href {http://adsabs.harvard.edu/abs/2010A%26A...522A...1K} {522, A1}

\bibitem[\protect\citeauthoryear{{Kissin} \& {Thompson}}{{Kissin} \&
  {Thompson}}{2015}]{kissin:15}
{Kissin} Y.,  {Thompson} C.,  2015, \mn@doi [\apj]
  {10.1088/0004-637X/808/1/35}, \href
  {http://adsabs.harvard.edu/abs/2015ApJ...808...35K} {808, 35}

\bibitem[\protect\citeauthoryear{{Mosser} et~al.,}{{Mosser}
  et~al.}{2010}]{mosser:10}
{Mosser} B.,  et~al., 2010, \mn@doi [\aap] {10.1051/0004-6361/201014036}, \href
  {http://adsabs.harvard.edu/abs/2010A%26A...517A..22M} {517, A22}

\bibitem[\protect\citeauthoryear{{Mosser} et~al.,}{{Mosser}
  et~al.}{2012}]{mosser:12}
{Mosser} B.,  et~al., 2012, \mn@doi [\aap] {10.1051/0004-6361/201220106}, \href
  {http://adsabs.harvard.edu/abs/2012A%26A...548A..10M} {548, A10}

\bibitem[\protect\citeauthoryear{{Paxton} et~al.,}{{Paxton}
  et~al.}{2013}]{paxton:13}
{Paxton} B.,  et~al., 2013, \mn@doi [\apjs] {10.1088/0067-0049/208/1/4}, \href
  {http://adsabs.harvard.edu/abs/2013ApJS..208....4P} {208, 4}

\bibitem[\protect\citeauthoryear{{Townsend} \& {Teitler}}{{Townsend} \&
  {Teitler}}{2013}]{townsend:13}
{Townsend} R.~H.~D.,  {Teitler} S.~A.,  2013, \mn@doi [\mnras]
  {10.1093/mnras/stt1533}, \href
  {http://adsabs.harvard.edu/abs/2013MNRAS.435.3406T} {435, 3406}

\makeatother
\end{thebibliography}







\bsp	
\label{lastpage}
\end{document}